# Macro pose based non-invasive thermal comfort perception for energy efficiency

Bin Yang[1,3], Xiaogang Cheng[2,4,5,*], Dengxin Dai[4], Thomas Olofsson[3], Haibo Li[5], Alan Meier[6]

[1] School of Environmental and Municipal Engineering, Xi'an University of Architecture and Technology, Xi'an, 710055, China
[2] College of Telecommunications and Information Engineering, Nanjing University of Posts and Telecommunications, Nanjing, 210003, China
[3] Department of Applied Physics and Electronics, Umeå University, 90187 Umeå, Sweden;
[4] Computer Vision Laboratory, Swiss Federal Institute of Technology (ETH), Zürich, 8092, Switzerland
[5] School of Electrical Engineering and Computer Science, Royal Institute of Technology, Stockholm, 10044, Sweden
[6] Lawrence Berkeley National Laboratory, Berkeley, 94720, USA

Corresponding author: Xiaogang Cheng (chengxg@njupt.edu.cn)

This study was supported by the National Natural Science Foundation of China (No. 61401236), the Jiangsu Postdoctoral Science Foundation (No. 1601039B), the Key Research Project of Jiangsu Science and Technology Department (No. BE2016001-3).

**Abstract:**

   Individual thermal comfort perception gives important feedback signals for energy efficient control of smart buildings. However, there is no effective method to measure real-time thermal comfort status of individual occupant until now. For overcoming this challenge, a novel macro posed based non-invasive perception method for thermal comfort (NIMAP) was presented. The occupant pose images were captured by normal phone camera (computer or cell phone) and the corresponding 2D coordinates can be obtained. Based on this, a novel pose recognition algorithm for thermal comfort, including 12 sub-algorithms, was presented. The 12 thermal comfort related macro poses can be recognized. Further, based on Fanger's theory, 369 subjects were invited for subjective questionnaire survey. 3 human occupants participated in the validation of the proposed method and massive data were collected. All the 12 thermal comfort related poses can be recognized effectively.

**Keywords:** Non-invasive measurement, Thermal comfort, Macro pose, Energy efficiency, Artificial intelligence

## 1. Introduction

   About 21% of global energy consumption occurs in residential and commercial buildings, and roughly half of that is consumed by their heating, ventilation and air conditioning (HVAC) systems in order to provide thermal comfort to their occupants [1-2]. However, much of this energy is wasted through overheating and overcooling beyond thermal comfort requirements. Real-time and accurate measurements of occupant thermal comfort can give feedback signals and reduce the energy consumption of HVAC. Many methods for evaluating thermal comfort have been proposed, including questionnaires, environment measurement, invasive physiological measurement and semi-invasive physiological measurement. Recently, non-invasive evaluations of thermal comfort have been proposed. Ghahramani used an infrared sensor on the frame of glasses mounted to measure skin temperature [3, 4], and Meier employed a Kinect to observe gestures associated with thermal discomfort [5]. These methods have significant practical limitations. The Kinect is unique device used for computer gaming with exclusive patent rights. Not everybody can be expected to wears glasses, so that the broad adoption of these methods is limited [3-5].

   Regardless of the approach, all methods must take into account that thermal comfort is an individual, subjective feeling that constantly changes as the person interacts with the surrounding environment [6, 7]. Until now, there has been no effective method to track human thermal comfort in practical applications, that is, outside of carefully controlled laboratory environments. To overcome these drawbacks, a new, non-invasive measuring method for measuring human thermal comfort is proposed. The method relies on poses obtained from the OpenPose model. A pose can be captured by a normal computer or cell phone camera and is then digitized. We defined 12 macro-poses that reflect human thermal preferences. The OpenPose data were combined with thermal comfort perceptions obtained through questionnaires, and a novel macro-pose recognition algorithm is constructed, including 12 sub-algorithms. The 12 macro-poses can be estimated. 3 human subjects were invited for algorithm validation, and a big dataset was collected.

   The main contributions of this paper are:
(1) 12 macro-poses of thermal comfort were defined based on 369 questionnaires of thermal comfort.
(2) A method for estimating human thermal preference based on data collected from normal standard computers or phone cameras was demonstrated.
(3) A novel algorithm (NIMAP) based on the OpenPose was developed for real-time measurement of human thermal preference.

   The rest of this paper is organized as follows. In Section 2, the related work about thermal comfort measurements is introduced. In Section 3, the research method, including macro-pose definition and OpenPose platform, are introduced. Based on this, the micro-pose recognition algorithm is proposed. Data validation results and discussion are shown in Section 4 and 5. Finally, conclusions are given in Section 6.

## 2. Related work

Human thermal comfort is subjective feeling which involves human psychology and human environment interaction [6]. Fanger has been explored thermal comfort since 1970s and founded Fanger's theory [6]. Based on it, many researchers study this topic in the past several decades.

A questionnaire is a useful and human-centered method for capturing subjective evaluation from occupants [8]. However, it is not feasible for occupants to give their real-time feedback [9]. For solving the practical goal of maintaining acceptable thermal comfort, the building industry adopted a method relying on measurement of environmental parameters, including indoor temperature, humidity and airflow. From the perspective of constant indoor parameters, ASHRAE and ISO (No. 7730) defined thermal comfort environment which is 'at least 80% of building occupants are psychologically satisfied with the temperature range of thermal environment' [10, 11]. For overcoming this problem, based on Back Propagation (BP) neural network, Liu [12] studied the individual thermal comfort and constructed a neural network-based model. Afroz [13] proposed a nonlinear autoregressive network to predict indoor temperature, and the network size was tuned for improving the efficiency of prediction model. However, the inter- and intra-individual differences existed in human thermal comfort [7]. Different people has different feeling in the same indoor environment. Therefore, physiological measurement method was explored by many researchers, including invasive measuring method, semi-invasive measuring method and non-invasive measuring method.

Skin temperature is an intermediate variable usually used for human thermal comfort estimation. Wang [14] studied the relationship between human thermal sensation and upper-extremity skin temperature. Nakayama [15] estimated human thermal sensation based peripheral skin temperature, and a subjective experiment was performed to analyze the relationship between peripheral skin temperature and subjective sensation votes. Liu [16], Takada [17], Sim [18], Wu [19] and Chaudhuri [20] proposed all kinds of methods to predict thermal comfort based on skin temperature. Heart rate variation (HRV) and electroencephalograph (EEG) were also explored. Yao [21] explored HRV and EEG to estimate thermal comfort. The results show that HRV and EEG can be factors to reflect human thermal comfort. Further, machine learning method was introduced into invasive physiological measurement method. Chaudhuri [22] proposed a data-driven method and 3 thermal comfort levels were defined which are cool-discomfort, comfort and warm discomfort. Based on this, support vector machine (SVM), artificial neural network (ANN), and logistic regression (LR), etc were used for constructing classifiers. Dai [23] combined machine learning with skin temperature, and an intelligent control method based on SVM was proposed. The validation results show that 3 skin sampling points can produce enough information for estimating thermal comfort, and the SVM classifier with linear kernel is better than that with Gaussian kernel. Kim [24] proposed a personal comfort models for predicting occupant thermal sensation. The data was collected from a personal comfort system (PCS) chair, and machine learning was used for data analysis.

Semi-invasive measuring method was presented in study [3, 4]. Ghahramani [3] collected skin temperature from 3 sampling points around human eyes. The infrared sensors are constructed on eye glasses, and some subjects are invited for subjective experiments. Further, Ghahramani [4] used unsupervised learning method to further analyze the data collected in study [3], and proposed a hidden Markov model to estimate skin temperature and thermal comfort.

The disadvantages of invasive measuring method and semi-invasive measuring method are obvious. The close-fitting sensor is required to collect human physiological parameters, which impedes widely practical application. As a solution, Cheng [25] used normal computer and cell phone camera to predict human thermal comfort. Two saturation-temperature (ST) models were proposed, including non-invasive thermal comfort measurement based on ST model (NIST) and non-invasive thermal comfort measurement based on partly ST model (NISPT). In the study [25], subtleness magnification technology was adopted and the Euler Video Magnification (EVM) was combined with big data to magnify skin features. The study [25] is a meaningful attempt in non-invasive thermal comfort perception.

In recent years, the deep learning techniques have been applied to this problem [26]. Besides the study [22-24], many other researchers made some meaningful attempts in the combination of machine learning and thermal comfort prediction. SVM was mainly used [27-29], and public dataset was used for method validation. Peng [30] used unsupervised and supervised learning to predict occupant behavior, and a demand-driven method was presented which was validated in 11 rooms of a commercial building.

Recently, non-invasive evaluations of thermal comfort have been proposed. XXX used an infrared sensor on the frame of glasses mounted to measure skin temperature [3, 4. Meier et al. employed a Kinect to observe gestures associated with thermal discomfort [5]. They defined 4 poses associated with thermal discomfort, established rule-based relationships, and detected them with the Kinect They also proposed a "library of thermal comfort gestures" to store the pose information. These methods have significant practical limitations. The Kinect is unique device typically used for computer gaming and many exclusive patent rights. For the infrared sensor mounted on glasses, not everybody can be expected to wears glasses. The broad adoption of these methods is therefore limited.

From the perspective of practical application, non-invasive measuring method of thermal comfort is the research direction in building industry. Further, human posture can reflect human thermal comfort status. With the development of deep learning technology, OpenPose was proposed and it is a deep learning-based open source platform [31-33]. OpenPose can produce key point coordinate of human skeleton and will be useful to estimate human thermal comfort. Based on OpenPose, we proposed a thermal comfort perception method (NIMAP) which belongs to non-invasive physiological measurement method.

## 3. Research method

### 3.1 Macro-pose definition and questionnaire

Human will naturally have various postures when they feel hot or cold. Thereby, based on Fanger's theory of thermal comfort, we defined 12 macro-poses. As shown in Table 1, the macro-poses are wiping sweat, fanning with hands, shaking T-shirt, scratch head, roll up sleeves, walking, shoulder shaking, folded arm, leg cross, hands around neck, warm hands with breath and stamping feet. Further, the thermal comfort level of wiping sweat and fanning with hands are 'hot', and the corresponding score is '3'. The other 10 poses have different thermal comfort level and score too. Fig. 1 and Fig. 2 show the continuous changes of 12 macro-poses in the time dimension. Sitting work style and standing work style are both considered in this paper, so that a total of 4 sub-fig were shown in Fig. 1-2.

In order to assess whether the macro-poses defined in this paper are in line with the human thermal sensation, a subjective questionnaire was used. All the subjects of questionnaire were required to assess the 12 macro-poses from the following options: (1) is it a cold action response, (2) is it a heat action response, or (3) neither. In addition, personal information of subjects, including height, weight, gender and age, was collected. Based on the results of questionnaire, the macro-poses were used for algorithm design and validation.

### 3.2 Algorithm

The human body naturally adjusts its position to maximize thermal comfort. These movements of bones and joints will produce various changes in space. Let $sp_i$ denotes the key points of human skeleton, since the images captured by normal camera is 2D, so that $sp_i$ is

$$sp_i = [x_i, y_i], i = 1, \ldots, k \tag{1}$$

where, $x_i$ and $y_i$ denote the horizontal and vertical coordinates in image space. The '$i$' denotes different key point of human skeleton and its number. The '$k$' is the maximum of key point number. If $sp_i$ is obtained accurately, the corresponding algorithm can be constructed for computing the movements of human body and recognizing human poses.

OpenPose is a kind of convolutional pose machines which can provide a sequential prediction framework for learning rich implicit spatial models [31-33]. Based on this, a human skeleton can be obtained. As shown in Fig. 3, there are 19 key points ($k = 19$). The key points of skeleton are nose, neck, right shoulder, right elbow, right wrist, left shoulder, left elbow, left wrist, right hip, right knee, right ankle, left hip, left knee, left ankle, right eye, left eye, right ear and left ear. The corresponding number is from 0 to 17. It should be noted that '$i=18$' denotes the scene background.

The framework of NIMAP algorithm presented in this paper is shown in Fig. 4. The images of occupant, work in office, is captured by normal computer camera. After processing by 'preprocessing' module and 'OpenPose' module, the key points coordinates with confidence value ($z$) will outputted. If the confidence value is less than a threshold ($z < \varepsilon$), the images will be discarded in this paper. If $z \geq \varepsilon$, the images will be imputed into poses recognition module. Finally, the type of macro-poses and its corresponding thermal comfort level can be obtained.

For poses recognition, we designed a sub-algorithm for each pose estimation. Due to the need of technical processing, 'walk' and 'stamping feet' belong to the same sub-algorithm. Further, 'hands around neck' and 'warm hands with breath' belong to the same sub-algorithm. Therefore, a total of 10 sub-algorithms were constructed for different macro-poses in this paper.

The distance ($L$) between the skeleton key points is European distance. For convenience of calculation, a standard distance is defined in this paper, that is

$$L_s = |sp_7 - sp_6| \qquad (2)$$

where $sp_7$ denotes the left wrist and $sp_6$ denotes the left elbow. Based on formula (2), the relative distance between the key points can be computed as

$$L_r = \frac{L_s}{L} \qquad (3)$$

different relative distance ($L_r$) will be computed, and different threshold of $L_r$ will be set for different poses recognition. Further, $L_{r\_max}$ and $L_{r\_min}$ will be set for some poses recognition which denote the extremum of $L_r$. In addition, mathematical slope is adopted in our algorithm. Further, coordinate changes of key points in continuous image frames are also used for pose estimation. The details of NIMAP algorithm is shown in Table 1.

**TABLE 1.** Pose definition based on Fanger's seven point scale

| No. | Pose category | Score | Thermal comfort Level | Can NIMAP recognize the pose or not |
|---|---|---|---|---|
| 1 | Wiping sweat | 3 | Hot | √ |
| 2 | Fanning with hands | 3 | Hot | √ |
| 3 | Shaking T-shirt | 2 | Warm | √ |
| 4 | Scratch head | 2 | Warm | √ |
| 5 | Roll up sleeves | 1 | Slight warm | √ |
| 6 | Walking | 0 | Neutral | √ |
| 7 | Shoulder shaking | -1 | Slight cool | √ |
| 8 | Folded arm | -2 | Cool | √ |
| 9 | Leg cross | -2 | Cool | √ |
| 10 | Hands around neck | -2 | Cool | √ |
| 11 | Warm hands with breath | -3 | Cold | √ |
| 12 | Stamping feet | -3 | Cold | √ |

**TABLE 2.** Macro pose-based non-invasive thermal comfort perception for energy efficiency

| Algorithm: NIMAP |
|---|
| **Output:** Pose category, thermal comfort level ([-3, 3]), thermal preference (-1, 0, 1) |
| **Step:** |
| 1. Surveillance video preprocessing |
|     (1) Frame extraction. |
|     (2) De-noise. |
|     (3) Region of interest (ROI). |
| 2. Searching coordinates of key points |
|     (1) Calling OpenPose platform. |
|     (2) Generating Jason. |
|     (3) Saving valuable frames based on Confidence value ($\varepsilon$ =0.5). |
| 3. Generating coordinate matrix based on Jason (18 × 3). |
| 4. Computing standard distance ($\alpha$), define the nearest distance threshold ($\tau$ =1.5). |
| 5. Pose and thermal comfort estimation. |
|     (1) Computing relative Euclidean distance. |
|     (2) Computing slope. |
|     (3) Computing movement speed. |
|     (4) A total of 10 sub-algorithms were constructed and called for 12 macro-poses. The 'walking' and 'stamping feet' belong to the same sub-algorithm. The 'hands around neck' and 'warm hands with breath' belong to the same sub-algorithm. |
|     (5) The sub-algorithms are 1) wiping sweat, 2) fanning with hands, 3) shaking T-shirt, 4) scratch head, 5) roll up sleeves, 6) walking and stamping feet, 7) shoulder shaking, 8) folded arm, 9) leg cross, 10) hands around neck, warm hands with breath. |
|     (6) Some parameters set: 1) wiping sweat: $L_r=1.8$, 2) fanning with hands: $L_{r\_max}= 120$, $L_{r\_min}= 80$, 3) Shaking T-shirt: 1.8, 120 and 80 are all used. 4) scratch head: $L_r = 1.8$, 5) Roll up sleeves, $L_r = 0.9$, 6) walk: $L_r = 1.8$, 7) stamping feet: slope difference threshold is 30. 8) shoulder shaking: $L_r = 1.5$, 9) folded arm, $L_r = 2$, 10) leg cross: $L_r = 1$, 11) hands around neck and warm hands with breath: $L_r = 3$. |
|     (7) The key points used for calculating $L_r$ are different for different macro-poses. |
| 6. Optimize algorithm parameters. |

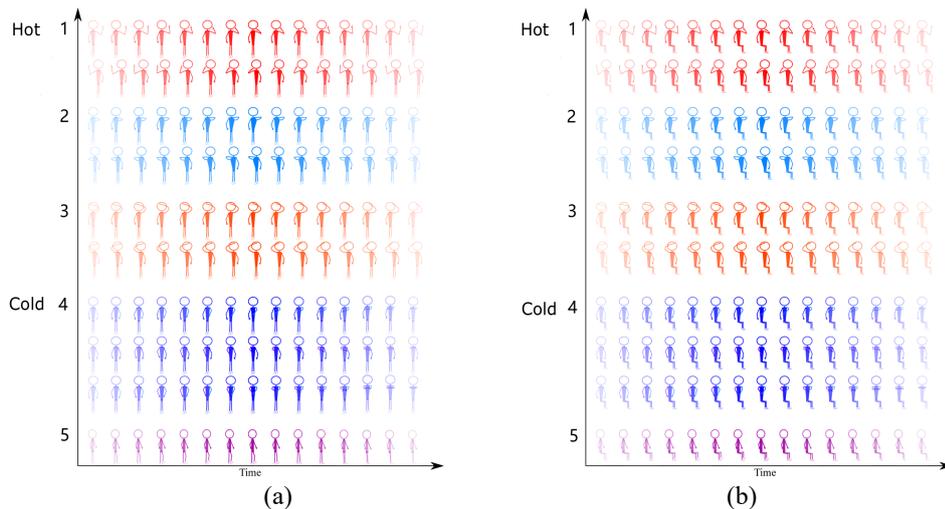

Fig. 1. Macro-pose of thermal comfort Part A ((1) a. stand work style b. seated work style. (2) in Fig. 1, '1' to '5' respectively denote fanning with hands, shaking T-shirt, wiping sweat, folded arms and stamping feet).

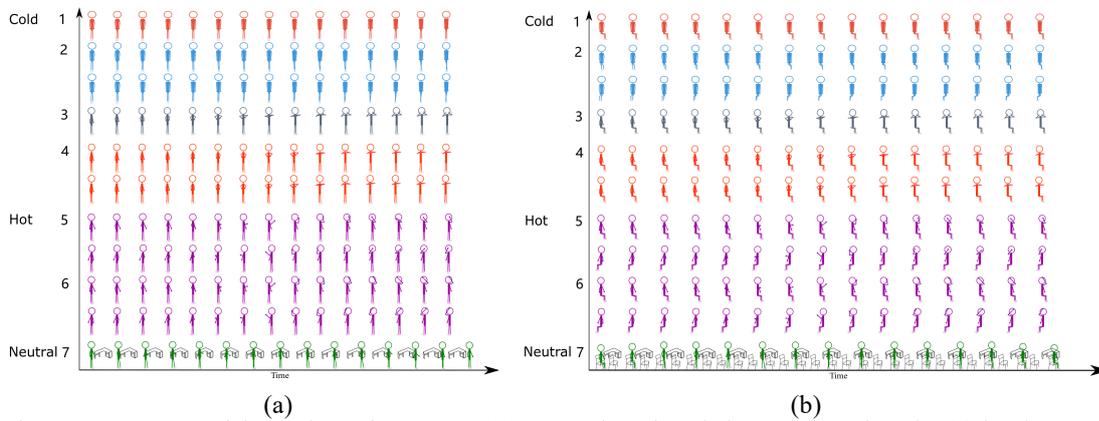

Fig. 2. Macro-pose of thermal comfort Part B ((1) a. stand work style b. seated work style. (2) in Fig. 2, '1' to '6' respectively denote shoulder shaking, leg cross, warm hands with breath, hands around neck, scratch the side of head, scratch the top of head and walking).

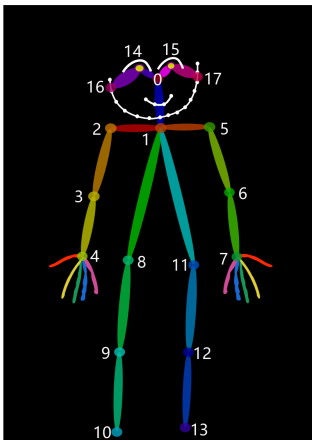

Fig. 3. Human skeleton and key points [31] ( 0. Nose 1. Neck 2. Right shoulder 3. Right elbow 4. Right wrist 5. Left shoulder 6. Left elbow 7. Left wrist 8. Right hip 9. Right knee 10. Right ankle 11. Left hip 12. Left knee 13. Left ankle 14. Right eye 15. Left eye 16. Right ear 17. Left ear 18. Background).

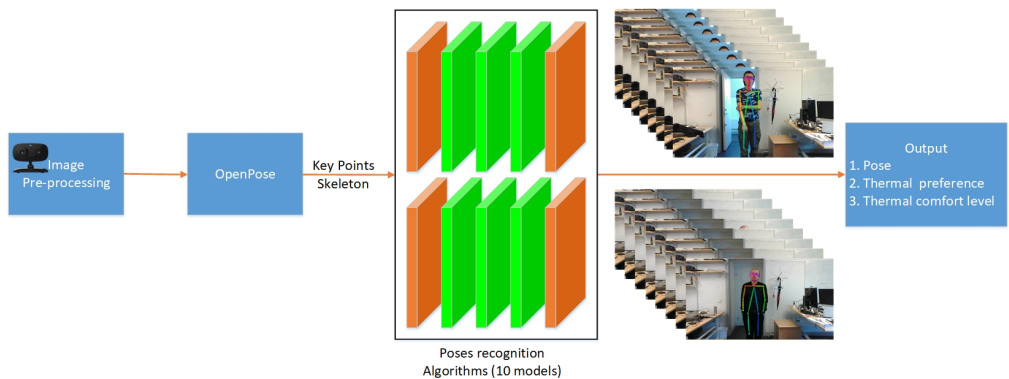

Fig. 4. The NIMAP network framework.

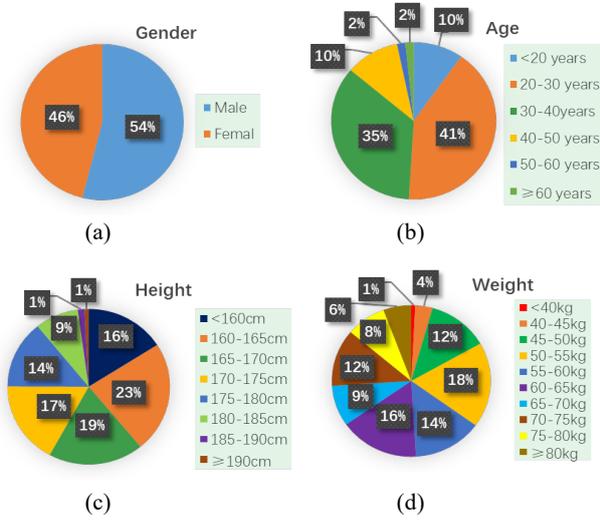

Fig. 5. Subjects information (Gender, age, height and weight).

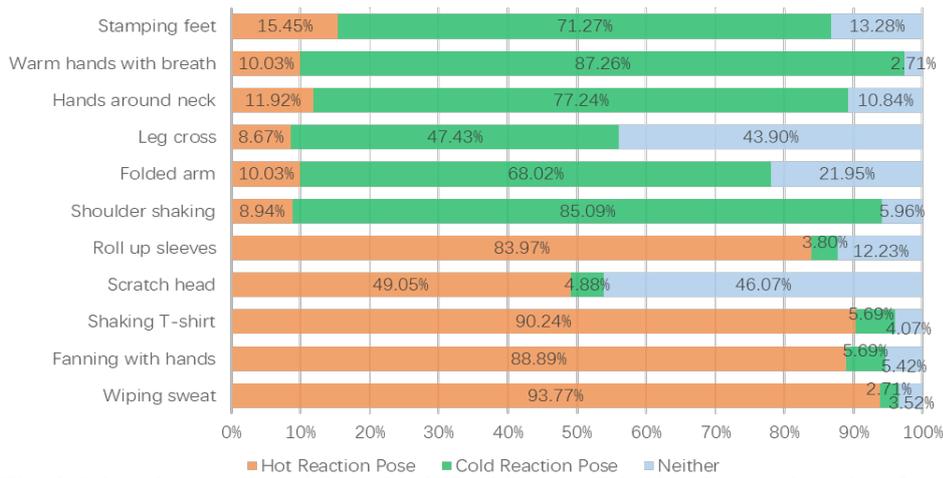

Fig. 6. Subjective questionnaire results (369 subjects participating the questionnaire of thermal comfort reaction).

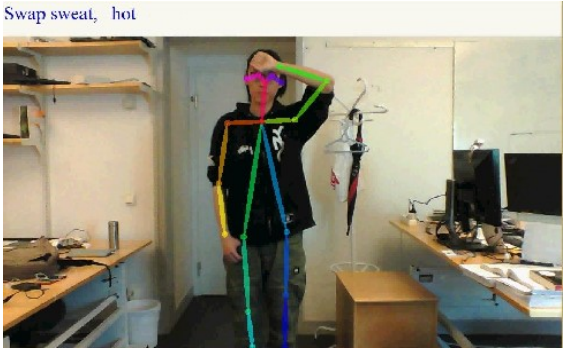

(a)

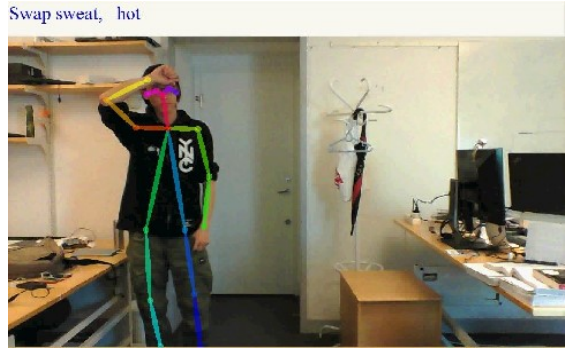

(b)

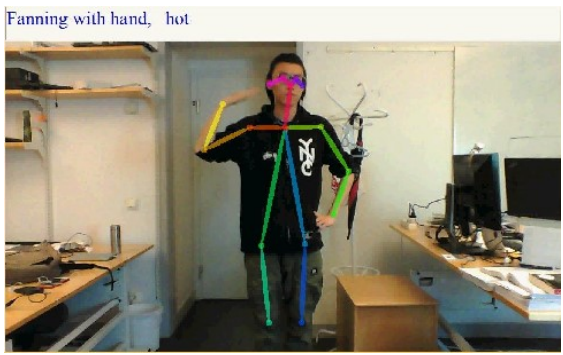

(c)

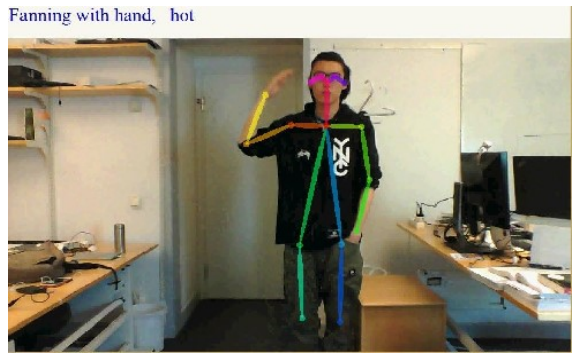

(d)

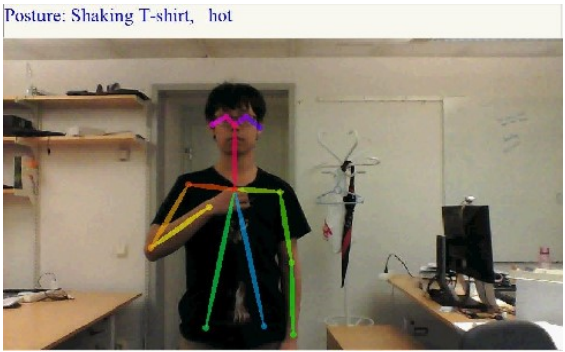

(e)

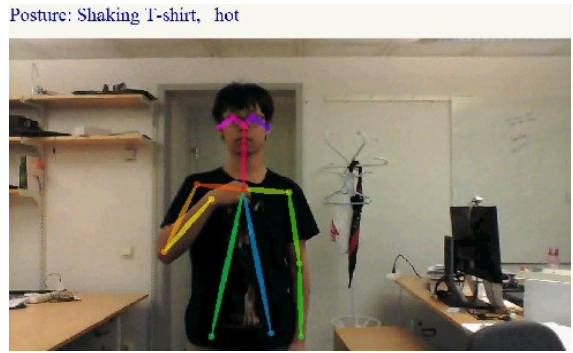

(f)

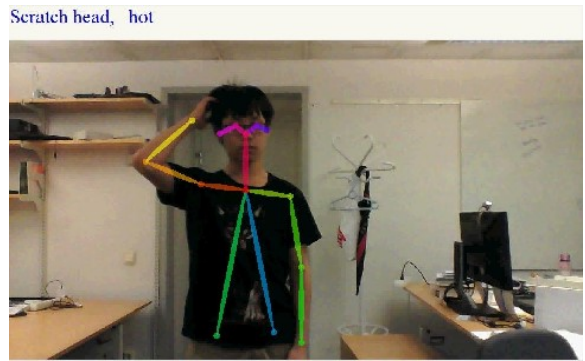
(g)

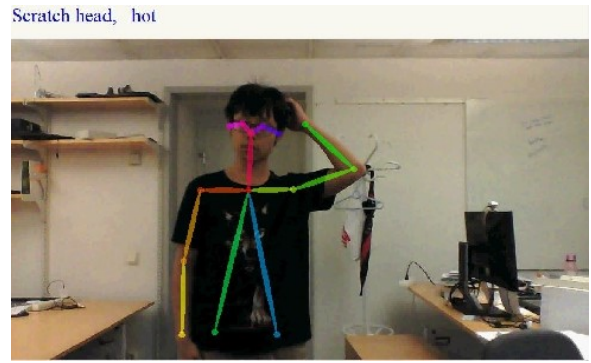
(h)

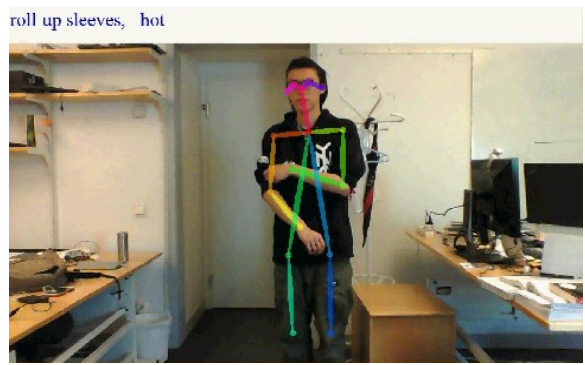
(i)

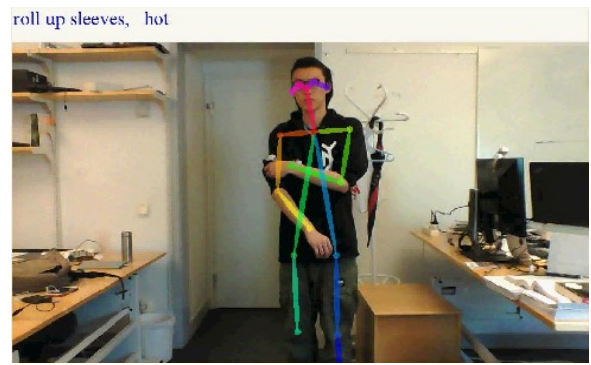
(j)

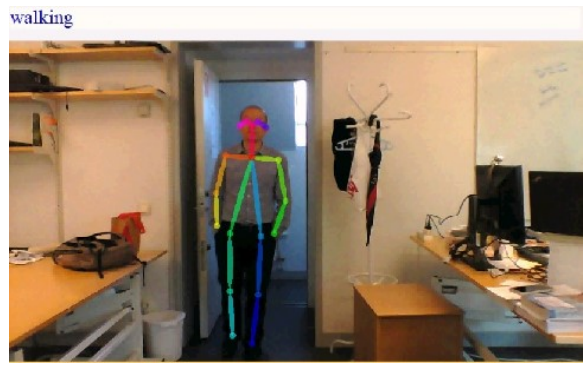
(k)

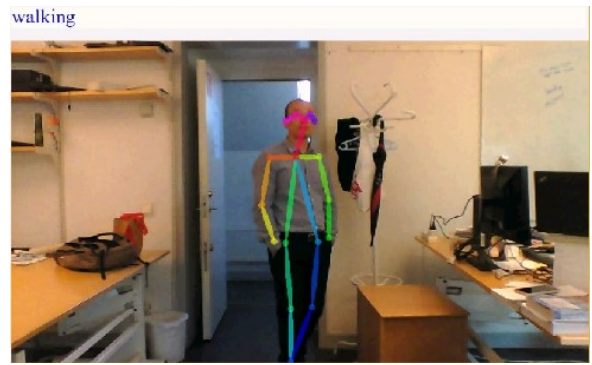
(l)

Fig. 7 Recognition results of hot and neutral poses by NIMAP algorithm proposed in this paper (Fig. 7.a-j denote the hot feeling poses which are wiping sweat, fanning with hands, shaking T-shirt, scratch head and roll up sleeves respectively. Fig. 7.k-I denote the neutral feeling poses which is walking.).

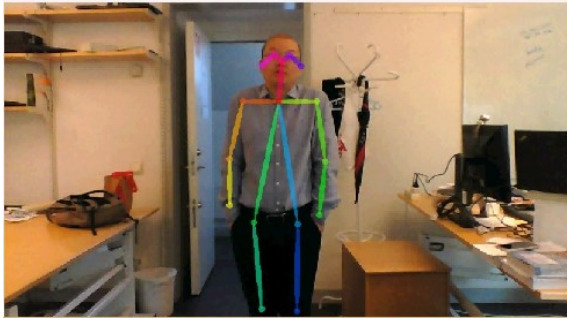
(a)

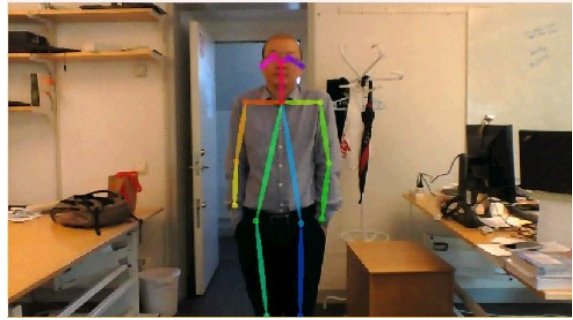
(b)

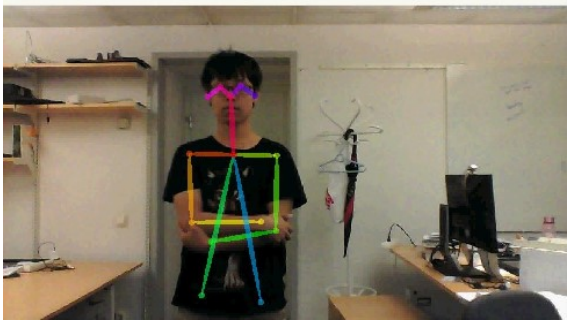
(c)

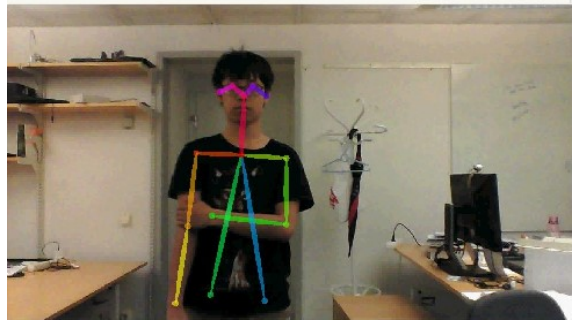
(d)

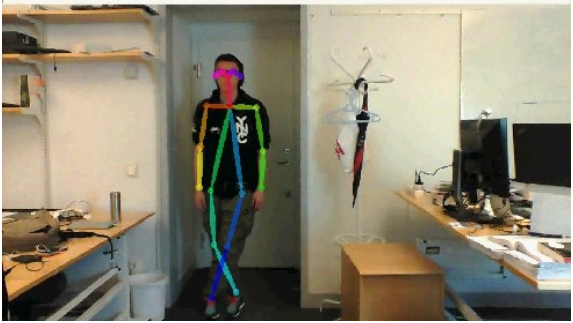
(e)

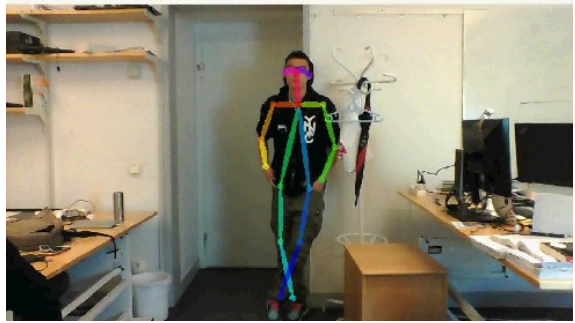
(f)
10

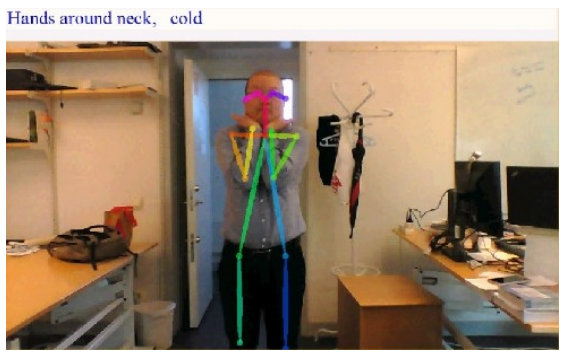
(g)
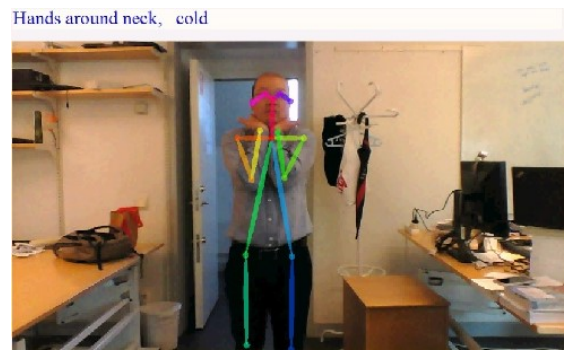
(h)
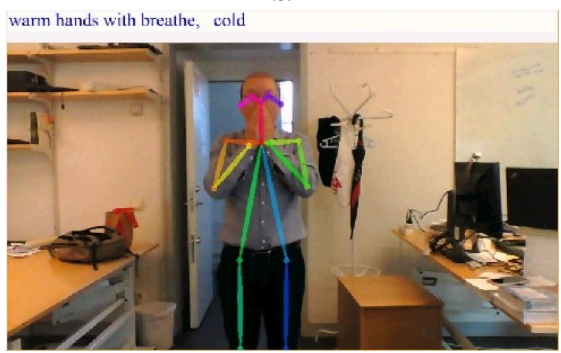
(i)
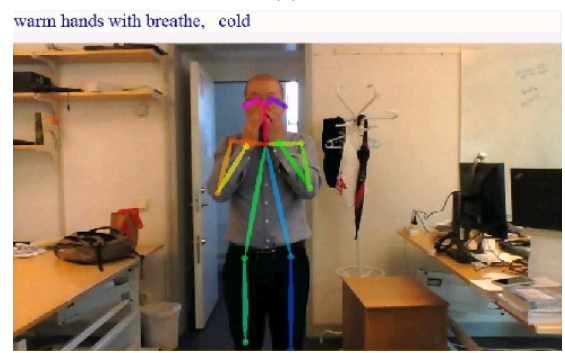
(j)
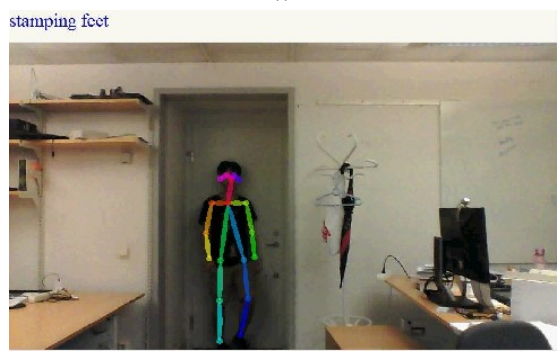
(k)
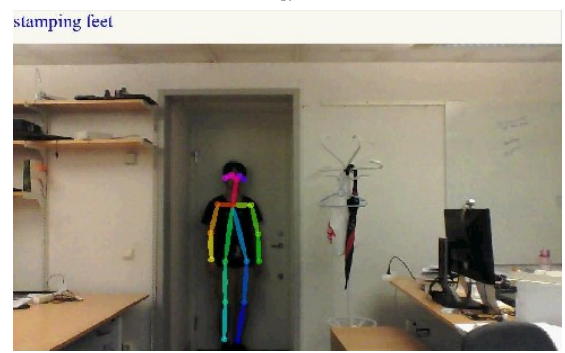
(l)

Fig. 8 Recognition results of cold feeling poses by NIMAP algorithm proposed in this paper (The cold feeling poses are shoulder shaking, folded arm, leg cross, hands around neck, warm hands with breathe and stamping feet, respectively).

## 4. Results

To validate the non-invasive thermal comfort perception method based on macro poses (NIMAP) presented in this paper, a subjective questionnaire was explored and a total of 369 valid questionnaire were collected for defining thermal comfort poses. Further, an algorithm validation experiment were handled, and 3 subjects are invited for real-time test.

In this paper, a computer workstation, with 64-bit version and 32G RAM, was used for algorithm validation. Graphics processing unit (GPU) is required in the process of algorithm training and test. The GPU adopted in this paper is NVIDIA GeForce GTX 980 (1920 × 1080, 32 bit, 60Hz), and the processor is Intel (R) Xeon (R) CPU E5-2687W V3 @ 3.10GHz.

As shown in Fig. 5, in all the subjects who output valid questionnaires, male and female are 199 and 170, respectively, which is relative balance. The age of most of the subjects are between 20 years and 50 years. Among them, two age ranges which are [20, 30) and [30, 40) accounted for 35% and 41%, respectively. Subject height statistics were performed at intervals of 5 cm. Among all the height intervals, the numbers of subjects whose height in [160, 165), [165, 170), [170, 175), [175, 180), [180, 185), [185, 190) are 85, 70, 63, 52, 32 and 5, respectively. The intervals of weight is 5kg. Among all the weight intervals, the numbers of subjects whose weight in [40, 45), [45, 50), [50, 55), [55, 60), [60, 65), [65, 70), [70, 75), [75, 80) are 13, 46, 67, 52, 60, 32, 45 and 30.

Based on Fanger's theory of thermal comfort, we defined 12 macro-poses which are hot reaction pose, neutral pose and cold reaction pose. The poses are wiping sweat, fanning with hands, shaking T-shirt, scratch head, roll up sleeves, walking, shoulder shaking, folded arm, leg cross, hands around neck, warm hands with breath and stamping feet. In questionnaire, all the subjects are required to assess the sensation of the 12 macro-poses defined in this paper, and the assessment results are shown in Fig. 6. The subjective assessment results of 10 of 12 macro-poses are fully consistent with our expectations and that of the remaining 2 poses are partially consistent with expectations. Fig. 6 shows that the definition of macro-poses in this paper is reasonable and meets the thermal sensation of human daily life.

As shown in Fig. 4, in practical application, all the images captured by normal computer camera will be processing by NIMAP algorithm. For improving estimating accuracy, when one image was inputted into NIMAP algorithm, the 12 sub-algorithms will be traversed according to a certain priority. The estimation results are shown in Fig. 7-8. A total of 3 subjects are invited for algorithm validation. Fig. 7 is the validation results for the hot and neutral poses. The pose order of validation results are the same as that in Table 2. From Fig. 7-a to Fig. 7-I, the poses are wiping sweat, fanning with hands, shaking T-shirt, scratch head, roll up sleeves and walking, respectively. Fig. 8 is the validation results of cold poses, the corresponding poses are shoulder shaking, folded arm, leg cross, hands around neck, warm hands with breathe and stamping feet, respectively. When the images were captured by normal computer or cell phone camera, the 2D coordinates can be obtained by OpenPose firstly. Further, the 12 sub-algorithms can recognize the poses based these coordinate point. In addition, the Euclidean distance and mathematical slope are adopted in this paper for poses recognition. The conditional thresholds of poses recognition used in this paper shown in Table 2.

## 5. Discussion

The large number of samples questionnaire handled in this paper show that the 12 macro poses defined is reasonable. Further, the validation results shown in Fig. 7-8 show that the 2D coordinates of the skeleton key points, as well as the displacement and slope changes of the coordinates, are very useful to the recognition of the thermal comfort poses.

Kinect were used for measuring human metabolism and thermal comfort which are meaningful attempt. However, the patent and price of Kinect make a limited practical application. In this paper, a normal computer camera or cell phone camera is required for the NIMAP proposed. Further, the NIMAP algorithm can be embedded into computer server of HVAC in building. Thereby it is convenient for practical application.

There are inter- and intra-individual differences in human thermal comfort. The NIMAP proposed is a kind of real-time thermal comfort perception method. The image frames (usually 30 or 24 frames/second) can be processed by NIMAP algorithm. Thereby, the real-time pose variation can be captured, and the intra-individual differences can be overcome. In the design and commissioning phase of the NIMAP algorithm, the relevant parameters were fine-tuned for different subjects, and a comprehensive parameter is given in this paper. In further applications, different parameters will be adopted according to population classification. So that the inter-individual differences can be overcome.

The study [25] is another meaningful attempt in non-invasive perception of thermal comfort. The differences between this paper and study [25] are shown as follow (1) the study [25] is from the perspective of skin color variation. However, human poses of sensation are used for feature extraction in this paper. (2) the output of the study [25] is predication values of skin temperature, and that of this paper is thermal comfort level and the type of human poses.

It should be noted that deep information of human body and office scene is not adopted in this paper. Signals interference with each other will happen, e.g. stamping feet and walk are sometimes misjudged during the debugging process. For overcoming this interference, more determine statements were used in the 12 sub-algorithms.

In questionnaire, the results of 2 of 12 poses defined in this paper, leg cross and scratch head, are just partly consistent with our expectation. For example, leg cross is a cold poses, but only 47.43% of the 369 subjects thought it is a cold poses, and 43.9% of the 369 subjects thought it is neither cold poses nor hot poses. There is similar results for scratch head. Thereby, some researcher maybe argue that why we still design the recognition algorithm for 'leg cross' and 'scratch head'? The reasons are as follows (1) it seems that there are description ambiguous for poses, leg cross and scratch head, in our questionnaire. So that the two poses were misunderstood by some subjects. Thereby, we still think that people will cross and stick their legs to warm when they feel cold. On the other hand, people will scratch side of head or top of head when they feel hot. (2) Even the two poses are not typical poses of thermal comfort, we would like to provide a kind of recognition method. So that it can be used in practical application if it is necessary.

6. Conclusion

The aim of this paper is to study a kind of non-invasive perception method of thermal comfort from macro-poses perspective. Based on Fanger's theory of thermal comfort, the subjective questionnaire was explored and the NIMAP algorithm was presented. The conclusion can be summarized as follows.
(1) The 12 macro-poses of thermal comfort defined in this paper can be used for describing human thermal sensation.
(2) The NIMAP algorithm presented in this paper is useful to recognize the 12 macro-poses and the human sensation level can be obtained.
(3) More abundant coordinate information of human body helps to perceive thermal poses accurately.

It should be noted that deep information of human body will be useful to perceive the human comfort sensation. It will be our work in next step.

7. Acknowledgement

The authors thanks to Yaser Sheikh's research group (Carnegie Mellon University) for providing his "Openpose" code.